\documentclass[fleqn,usenatbib]{mnras}

\usepackage[T1]{fontenc}

\DeclareRobustCommand{\VAN}[3]{#2}
\let\VANthebibliography\thebibliography
\def\thebibliography{\DeclareRobustCommand{\VAN}[3]{##3}\VANthebibliography}

\usepackage{graphicx}	
\usepackage{amsmath}	
\usepackage{amssymb}	

\usepackage{color, soul}  

\newcommand{\comment}[1]{}
\newcommand{\changed}[1]{{#1}}

\title[Astrosat:  Forecasting satellite transits]{Astrosat: Forecasting satellite transits for optical astronomical observations}

\author[J. Osborn et al.]{
James Osborn,$^{1}$\thanks{E-mail: james.osborn@durham.ac.uk}
Laurence Blacketer,$^{2}$
Matthew J. Townson,$^{1}$
and
Ollie J. D. Farley$^{1}$
\\
$^{1}$Centre for Advanced Instrumentation, Department of Physics, Durham University, DH1 3LE, UK\\
$^{2}$Northern Space and Security, Aykley Heads Business Centre, Aykley Heads, Durham, DH1 5TS, UK
}

\date{Accepted XXX. Received YYY; in original form ZZZ}

\pubyear{2021}

\begin{document}
\label{firstpage}
\pagerange{\pageref{firstpage}--\pageref{lastpage}}
\maketitle

\begin{abstract}
\comment{
200 words for Letters
}
The impact of large-scale constellations of satellites, is a concern for ground-based astronomers. In recent years there has been a significant increase in the number of satellites in Low-Earth Orbit and this trend is set to continue. The large number of satellites increases the probability that one will enter the field of view of a ground-based telescope at the right solar angle to appear bright enough that it can corrupt delicate measurements. 
We present a new tool ``Astrosat'' that will project satellite orbits onto the RA/DEC coordinate system for a given observer location and time and field of view. This enables observers to mitigate the effects of satellite trails through their images by either avoiding the intersection, post-processing using the information as a prior or shuttering the observation for the duration of the transit. 
We also provide some analysis on the apparent brightness of the largest of the constellations, Starlink, as seen by a typical observatory and as seen with the naked eye. We show that a naked eye observer can typically expect to see \changed{a maximum of} 5 Starlink satellites at astronomical twilight, when the sky is dark. With the intended 40000 satellites in the constellation, that number would increase to 30.

\end{abstract}




\section{Introduction}

The effect of the increasingly numerous satellite constellations on ground-based astronomy is an area of significant and timely concern for astronomers, see for example  \cite{McDowell2020,Hainaut2020,LSST20,Tyson2020,Williams2021,bassa2021analytical}. As a satellite transits through the field of view of the telescope an artefact can be imprinted into the scientific data, for example a bright streak across an image. In the worst case these artefacts can lead to the data being unusable. Therefore, there is a demand for new astronomical tools which will support astronomers in mitigating these detrimental effects. 

Several tools exist which can be used to find the position of a particular satellite. However, in order to mitigate the deleterious effects, astronomers need to know which satellites will intersect with a given part of the sky. Here we present a new Python tool, Astrosat\footnote{https://github.com/james-m-osborn/astrosat}, which calculates which satellites can be seen by a given observer in a given field of view at a given observation time and observation duration. This includes the geometry of the satellite and observer but we also estimate the expected apparent brightness of the satellite to aid astronomers in assessing the impact on their observations.

This tool could be used by several communities. Astronomers can use the tool to mitigate the effects of satellites on their scientific observations. This could be accomplished by scheduling observations to minimise the brightest satellite transits, developing active shuttering capabilities to block the short transit duration in real-time, or as a prior for post-processing techniques to identify the trails in raw data \citep{Hainaut2020}. Landscape and astro-photographers can likewise use the tool to plan their activities in order to avoid or even include the satellite trails depending on their desired impact. The tool could also be used for outreach activities to visualise and demonstrate the effect of the satellites on scientific observations but also how our naked-eye view of the night-sky is being impacted \citep{Venkatesan2020}. Astrosat is based on a central API which can be easily scripted into stand-alone tools or integrated into other software pipelines. 

\section{Astrosat}

\subsection{Orbit modelling}

The space object orbits are generated using the Simple General Perturbations 4 (SGP4) orbital propagator together with Two Line Element (TLE) sets acquired from Celestrak \footnote{https://celestrak.com/}. Here, we use TLEs for all active satellites (4633 satellites at the time of writing) and for the brightest objects (164 objects), including for example, rocket bodies. 

A TLE, which is two lines of 69 characters describing the orbit of a space object at a particular instant in time, is generated by fitting an SGP4-generated orbit to the result of an orbit determination process applied to a series of observations of a space object. The best-fitting orbital elements, and the epoch at which they are valid, are then encoded into the two lines of the TLE \citep{STR3}.

Because SGP4 is an analytical technique that uses simplified force modelling, and the numerical precision is limited to two lines of 69 characters, the positional accuracy of a TLE is also limited. However, it is not possible to quantify the accuracy of any individual TLE as it depends on the accuracy of the underlying observations, which are unknown. The positional uncertainty in a TLE of a large well-tracked space object is thought to be on the order of a few km, primarily in the in-track direction \citep{Vallado2012}. This uncertainty increases further as the position is propagated away from TLE epoch using SGP4.

The accuracy requirements for assessing the time at which a space object will transit the field of view will depend on the size of the field, the length of the exposure and the apparent angular velocity of the object. 

\changed{
If the TLEs are up to date (they are updated several times a day but the exact update rate of celestrak is unknown), then a worst-case estimate can be found by converting the expected orbital error into an angular or temporal error. 

For the in-track error, it is the associated temporal error that is important. 
The object will still pass through the field but at a different time than expected. 
The worst case is a low-altitude orbit viewed at zenith when the space object will have the largest apparent angular velocity. 
Assuming circular orbits, the apparent angular velocity can be calculated using,
\begin{equation}
    v = \frac{\sqrt{GM_E/(r+R_E)}}{z},
\end{equation}
where $G$ is the gravitational constant, $M_E$ is the mass of the Earth, $R_E$ is the radius of the Earth, $r$ is the altitude of the object and $z$ is the range to the object from the observer.

If we assume an object orbiting in a low orbit at 335~km, then a 2~km in-track error \citep{Vallado2012} will convert to 0.35~degree or 0.27~seconds. 
This will reduce for higher-altitude orbits and lower elevations but increase significantly for out of date TLEs. 
For example an in-track error of 25~km \citep{Vallado2012} will lead to an angular offset of 4.3~degrees or 3.26~seconds. 
To mitigate this error we list satellites with an expected intersection at +/- 5~seconds of the observation time such that all objects are listed. 
Therefore, due to the precision of the TLEs, it should be noted that precise time estimates are not possible. 
Despite the large component of TLE uncertainty in the in-track direction, early TLEs can still provide useful information on the intersections between the field and a space object orbit, if not the precise time of transit.

The cross-track error is more critical as it determines whether the object will pass through the field or not. 
The expected precision in the cross-track direction is better with values of a few hundred metres generally quoted \citep{Vallado2012}. 
If we assume a cross-track error of 500~m then the expected angular error for a low altitude object (335~km) at zenith (worst case) is 0.09~degree (5.2~arcminutes). 
This is difficult to mitigate and will have a larger real effect on smaller fields. The field of view could be expanded by this distance and the results interpreted probabilistically.
}

It is recommended that the tool is run as close as possible to the planned observation to ensure the highest possible accuracy. \changed{The TLEs are archived so can be used for post-processing at a later date without any loss of precision.}

\subsection{Brightness modelling}
\label{sect:brightness}
Not all satellite transits impact astronomical observations. For the satellite transit to be an issue it needs to have a significant number of photons to affect the analysis of the astronomical target. 
Hence it is important to not only know details of when and where a transit occurs, it is also important to know the expected brightness of the transiting satellite.
It is also worth pointing out that all attempts at estimating the expected apparent brightness of a satellite require assumptions on the size, orientation and specularity of the satellite, all of which are often unknown.

The apparent brightness of the satellite can be approximated by \citep{McCue1971},
\begin{equation}
    m = -26.74-2.5\log_{10}\left(\frac{A\gamma f(\phi)}{z^2}\right) + x\chi,
    \label{eq:mag}
\end{equation}
where $A$ is the cross-sectional area, $\gamma$ is the albedo or reflectivity, $\phi$ is the solar phase angle, $f$ is a function that defines the fraction of reflected light based on the solar phase angle, $z$ is the range to the satellite from the observer, $x$ is the atmospheric absorption coefficient, 0.12 mag/airmass in the V-band \citep{Patat2011,Hainaut2020} and $\chi$ is the airmass. Generally, for elevation angles, $\epsilon$, greater than 10 degrees from the horizon a simple $\chi=1/\cos{(\pi/2-\epsilon)}$ can be used. However, for low elevation angles, which can be the case for satellites, we need to include the curvature of the Earth. \changed{Here we estimate the airmass directly from the range to the object and orbital altitude of the object.}

\comment{
Here we use the empirical model of \citep{Kasten1989},
\begin{equation}
    \chi = \left[\sin{\epsilon} + a\left(\epsilon \times\frac{180}{\pi} + b\right)^{-c}\right]^{-1},
\end{equation}
where $\epsilon$ is in radians, a = 0.15, b=3.885 and c = 1.253.}

We can assume a diffuse spherical model for the satellite, in which case the function, $f$ is given by \citep{McCue1971},
\begin{equation}
    f(\phi)_\mathrm{diff\_sphere} = \frac{2}{3\pi}\left(\left(\pi-\phi\right)\cos{\phi}+\sin{\phi}\right).
    \label{eq:f}
\end{equation}
\changed{In recent publications, it has been demonstrated that Starlink satellites show very little dependence on the solar phase angle (see for example \cite{Horiuchi2020}). In Astrosat it is possible to use a diffuse spherical model or to set $f(\phi)=1$. In this study we use $f(\phi)=1$ for starlink satellites.}

This is obviously a very rough approximation. We need to assume a cross-sectional area and reflectivity coefficient for each satellite. As these are not known, we assume a circular cross-section with a radius of 1~m and $\gamma$=0.25. These values are thought to be representative of the Starlink-like satellites \citep{Hainaut2020}.

\changed{ 
Recent measurements of the new generation of Starlink visorsat satellites suggest that the V-magnitude has been reduced by 1.3~magnitudes \citep{mallama2021brightness}. Therefore, here we reduce the albedo by 30\% to reflect this.

In Astrosat the reflection cross-section can be provided by the user and therefore can be tuned as more data becomes available for Starlink and other satellites.
}

LEO constellations are likely to be the most disruptive to astronomical observations due to their low altitude and isotropic distribution. The exception to this will be rocket bodies which, although few in number, can be large and therefore appear very bright. With the growing interest in the space environment and satellite surveillance and tracking, we expect to be able to build a database for the $A\gamma$ product for different space object and satellite types.

Currently, this model is valid for the V-band. We intend to develop the model further to include a wavelength scaling and even consider the implications in the radio spectrum. This is left as an extension for the future.

The model does not include transient events, such as flares or rotating objects which are impossible to predict without a lot more information about a particular object and its dynamics.

\changed{
\subsection{Functionality}
Astrosat is based on a central application programming interface (API) which can be easily scripted into stand-alone tools or integrated into other software pipelines. 

As previously stated the space object TLEs are downloaded from Celestrak at run-time (if necessary). 
To validate the satellite positions we calculate the time and elevation angle at culmination for a random sample of 30 starlink satellites. That is the point of highest elevation angle as seen by an observer on the ground.

We compared the predicted satellite positions and time at culmination from Astrosat and Heavens-Above.com\footnote{https://www.heavens-above.com/}. We find that the elevation angle agrees to within the precision listed on Heavens-Above (1~degree) in every case. The time of culmination differs by 2.6 +/- 9.2~seconds. It is likely that the difference is due to the two systems using different TLEs.

The position of the stars is found by querying the SIMBAD Astronomical Database\footnote{http://simbad.u-strasbg.fr/simbad/}\citep{2000A&AS..143....9W} for fields less than 2~degrees or by searching the Yale Bright Star Catalogue\footnote{http://tdc-www.harvard.edu/catalogs/bsc5.html}\citep{hoffleit1991bright} for larger fields. 

To find the space objects that intersect with any given field of view, the software uses a nested search algorithm. 
Initially we estimate the apparent position of every object as seen by the observer every 10 seconds during the observation window. 
For all the satellites that come within 15~degrees of the target direction, we undergo a second more precise scan, probing the positions every 1~second. 
15~degrees is chosen as the fastest we expect an object to appear is 1.3~degrees/second for an object orbiting at 335~km (Starlink lower limit) at zenith. 
This ensures that even the fastest objects are seen in the search radius at least once with 10~second cadence.
We then interpolate between the one second position probes to estimate the exact times the object enters and exits the field of view. 
The location is checked at the mid-intersection time as a validation of the interpolation.
The solar phase angle, if used, can be calculated using the sine rule and the shadow of the Earth is also included.

To estimate the apparent position of the stars and space objects as seen by an observer we use pyephem\footnote{https://rhodesmill.org/pyephem/}. 
This is a well-used library and has been validated elsewhere. 
We have implemented previously published brightness estimation algorithms and these have also compared with measurements of Starlink satellites elsewhere (see section~\ref{sect:brightness}).

As an example of benchmark run-times, we probed 0.5~degree field-of-views as seen from an observer at Durham, UK, for 21:00 on the 27th September 2021 for a 1~hour observation. We include the current case of 1600 Starlink satellites as well as artificially generated TLEs for 12000 and 40000 Starlink satellites. The computer used for the benchmark tests is 2019 MacBook Pro, 2.8 GHz Quad-Core Intel Core i7 with 16GB memory. The tests took 18~seconds, 46~seconds and 3:20 minutes for the 1600, 12000 and 40000 satellite case respectively. These processing times are not prohibitive and could be improved with more sophisticated search algorithms.
}

\comment{
(Solar scaling, atmospheric absorption...)
}

\subsection{Vega Example}

\begin{figure}
    \centering
    \includegraphics[width=0.45\textwidth]{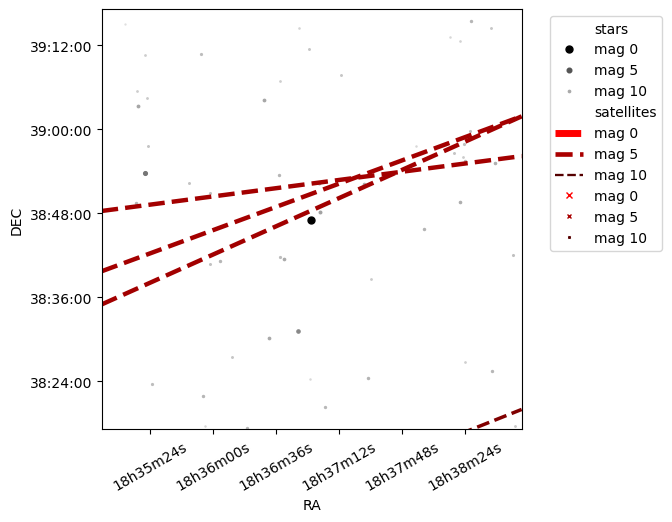}
    \caption{\changed{Example field image showing a 1~degree field of view around the star Vega. The brightness of the stars are shown by the grey scale and size of the point. The satellite trails are shown in red dashed line with line-width and red colour intensity denoting the apparent magnitude of the satellite. The designations, time of interception, duration of transit and expected brightness of the satellites are shown in the terminal, recreated in table~\ref{tab:vega_terminal}. This example is for 21:00 on the 27th September 2021 from Durham, UK.}}
    \label{fig:vega}
\end{figure}

Figure~\ref{fig:vega} shows an example output from the tool. In this example the observer was in Durham, UK, the target was set to Vega with a \changed{1~degree field at 21:00 on the 27th September 2021} for a one hour observation. The figure shows the bright star of Vega in the centre with other stars shown in black. The brightness of the star is shown by the size and brightness of the marker as indicated in the legend. The predicted satellite trails are shown in red dashed lines with the expected brightness designated by the thickness and saturation of the line, as shown in the legend. In this case four satellite intersections are predicted at magnitudes that can be seen on this scale. The designations, time of interception, duration of transit and expected brightness of the satellites are shown in the terminal, \changed{recreated in table~\ref{tab:vega_terminal}}.

\begin{table}
    \centering
    \begin{tabular}{l|c|c|c}
        Name & Time (UTC) & Duration (s) & Magnitude (V)  \\
STARLINK-2231    &              20:09:07  &      1.3      &      5.38\\
STARLINK-2719    &              20:26:13  &      1.2      &      5.42\\
GLOBALSTAR M080  &              20:32:48  &      0.4      &      7.46\\
STARLINK-1293    &              20:37:54  &      1.5      &      5.45\\
    \end{tabular}
    \caption{\changed{Example output for satellite intersection with a 1~degree field of view around the star Vega for 21:00 with a hour observation on the 27th September 2021 as seen from Durham, UK.}}
    \label{tab:vega_terminal}
\end{table}

\section{Analysis}

In this section we provide some analysis from Astrosat. We concentrate on Starlink satellites, as they are currently the most numerous constellation with plans to significantly expand. At the time of writing there are 1635 TLEs attributed to starlink satellites.
This means that they have the potential to have the greatest impact on an astronomical context.\comment{ In addition, previous studies have validated the brightness model for the Starlink satellites\hl{updateXXX} \citep{Hainaut2020}, this remains to be done for other satellite types.}

In addition to the apparent brightness of the satellite we also include the satellite density and apparent angular velocity all as a function of elevation angle. By combining these three parameters an observer can assess the effect of the satellites in terms of signal per pixel per exposure for a given system (pixel size and exposure time). \changed{This is important because it is the received flux per pixel per exposure that is important. For example a faint but slow satellite might have a great effect that a fast and bright satellite.}

Unless stated otherwise the analysis is presented for a `typical' observatory at 25~degrees North (or south) in Latitude and 1000~m altitude. The longitude is set to 0 but does not matter for this analysis. This typical observatory is used for this study as the latitude of the observatory does make a difference in the number of satellites visible above the horizon.
\comment{

Figure~\ref{fig:full_sky} shows a visualisation of the Starlink satellites (left) and all active satellites (right) projected onto an Right Ascension (RA) / Declination (DEC) projection with bright stars also shown in white. Any particular observer will be able to see a section of the RA/DEC space depending on latitude, longitude and local time. The satellites are not fixed in RA/DEC space and so the images shown are only for an example at 0 UTC on 16th July 2021.
}
\comment{
\begin{figure}
    \centering
    \includegraphics[trim=0 0.5cm 4cm 1cm,clip,height=7cm]{figures/Sky_field_full_sky_starlink.png}
    \comment{
    \includegraphics[trim=0 0.5cm 4cm 1cm,clip,height=6cm]{figures/Sky_field_full_sky_active.png}
    }
    \caption{Map of Starlink satellites projected onto RA and DEC coordinate system for midnight on 16th July 2021 for an observer at the centre of the Earth.}
    \label{fig:full_sky}
\end{figure}
}
\subsection{Number density of Starlink satellites}

\begin{figure}
    \centering
    \includegraphics[width=0.4\textwidth]{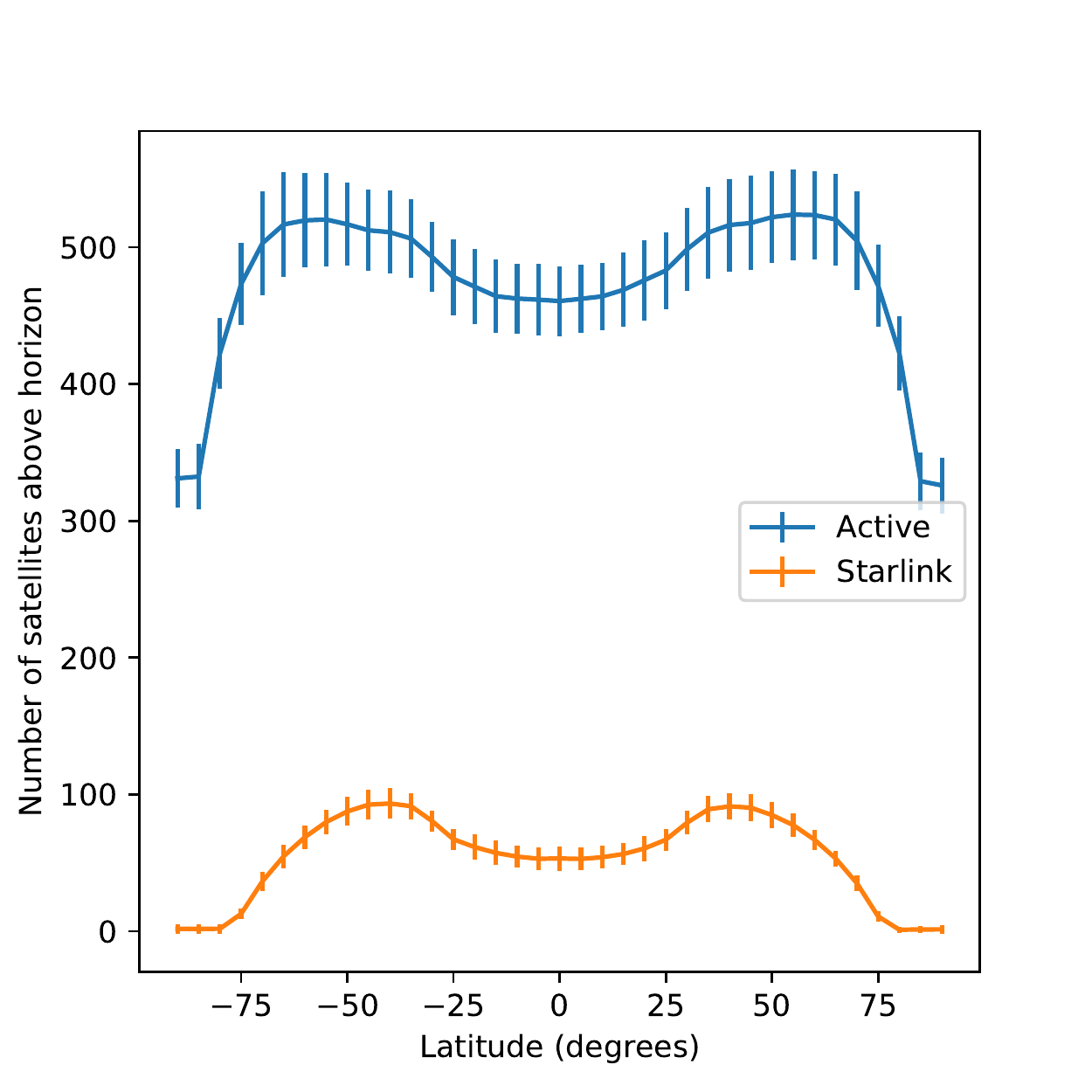}
    \caption{Number of satellites above horizon at any moment. 
    At the poles, only polar orbits are visible, hence the lower number of satellites visible. 
    The maximum is at approximately $\pm 50$~degrees caused by the increased density of mid-inclination orbits such as Starlink.
    \changed{
    To estimate the error bars we calculate the number of satellites visible above the horizon every hour for a 24-hour period. The error bars show the standard deviation of this value. 
    This is a useful tool to demonstrate the variance of the number of visible objects at any particular time.
    }
    \comment{
    , error bars are standard deviation over every hour (demonstrating variance with time).}
    }
    \label{fig:Nsat_lat}
\end{figure}

Figure~\ref{fig:Nsat_lat} shows the total number of satellites above the horizon as a function of latitude. Considering the line for all active satellites in blue, the plot shows that approximately 330 satellites are above the horizon to an observer located at either pole. These are the high numbers of systems that are in polar orbits. For an observer located at the equator, approximately an additional 120 satellites are visible, for a total of 450. This number now includes those satellites in lower inclination orbits that were not visible to an observer at the poles. The number of satellites above the horizon peaks at approximately $\pm 50$, which results in an increased spatial density caused by mid-inclination satellites such as Starlink. The shape of this plot agrees with the work of \cite{McDowell2020}.

\begin{figure}
    \centering
    \includegraphics[width=0.4\textwidth]{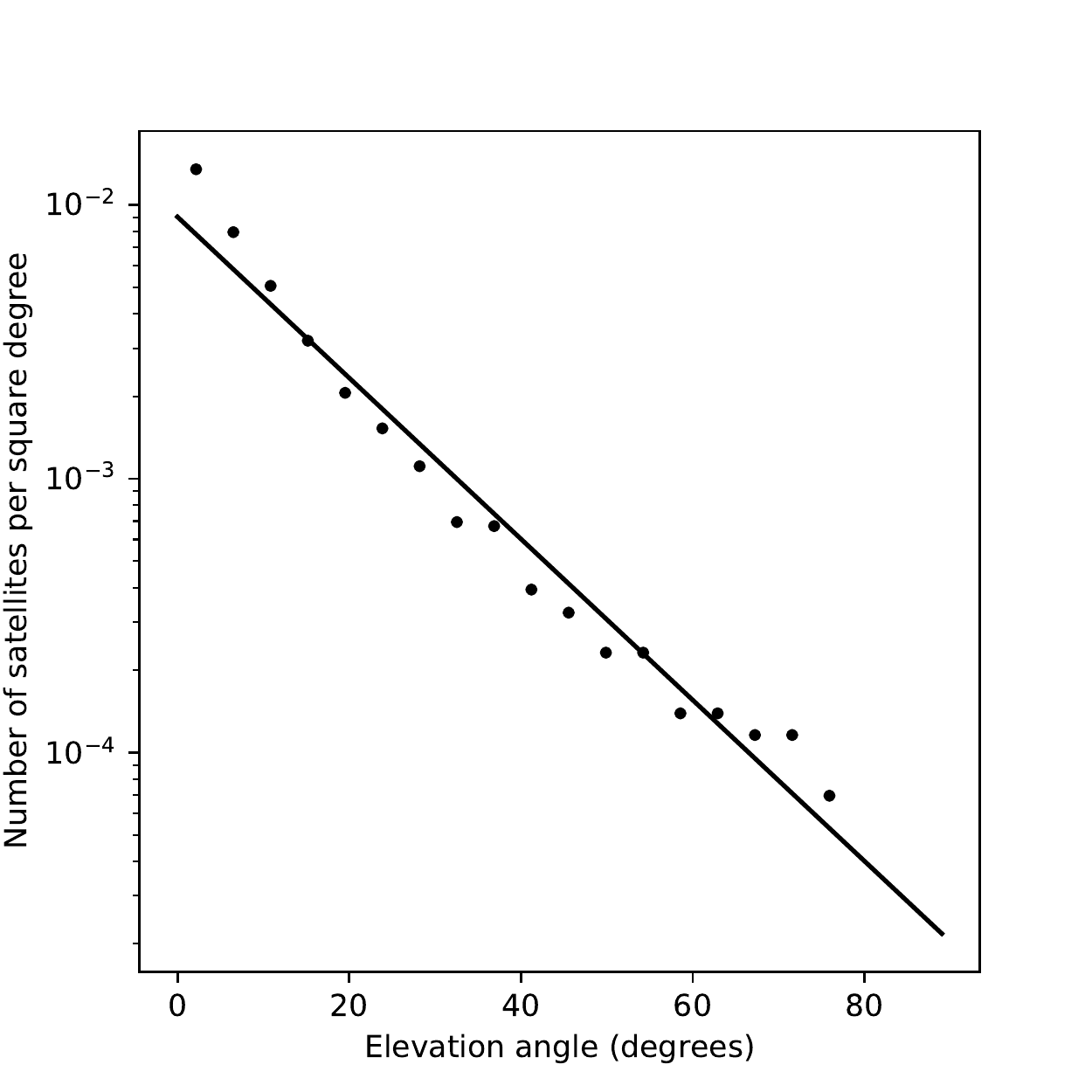}
    \caption{Number density of Starlink satellites as seen from a typical observatory at a latitude of +/- 25~degree. This is for a total number of 1635 Starlink satellites at the time of writing. The line of best fit is y = -0.03x-2.04.}
    \label{fig:satDensity}
\end{figure}

For astronomers, the total number of satellites in the sky is not so important. More interesting is the density of the satellites, or number per degree, and this varies as a function of elevation angle. Figure~\ref{fig:satDensity} shows that a higher density of satellites is seen near the horizon, as expected. \comment{We average over azimuth angle but it should be noted that the number of satellites will also have an azimuth dependence which will depend on the latitide of the observer.}

\subsection{Apparent brightness of Starlink satellites}

\begin{figure}
    \centering
    \includegraphics[width=0.4\textwidth]{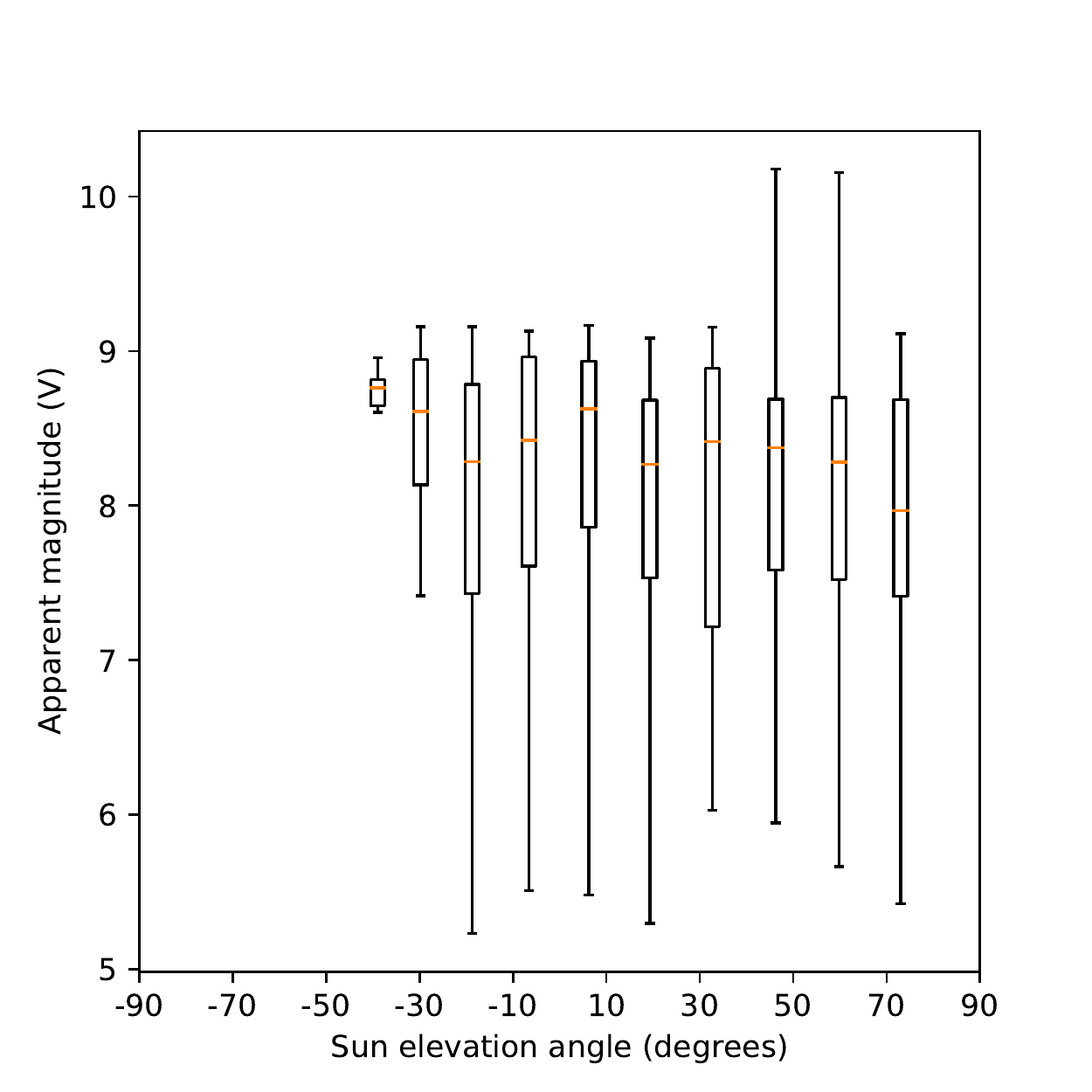}
    \caption{Apparent magnitude of Starlink satellites as a function of sun elevation (negative represents night-time when the sun is below the horizon) as seen by an observer at a typical observatory at a latitude of 25~degree North in the summer. We see that generally the Starlink satellites have an apparent magnitude of approximately \changed{8}. 
    As the sun elevation becomes more negative the \changed{brighter satellites are not seen. Brighter satellites tend to be in lower orbits and so are lost to the Earth's shadow sooner than the higher altitude, fainter satellites.}
    The large whiskers show the variability in the estimates which is mainly due to the variable satellite range.\comment{ and solar phase angle.}\comment{ We see that Visible satellites are mainly at lower elevation angles, where there is a higher density of satellites, but the brighter ones tend to appear at higher elevation angles.} \changed{The central horizontal line indicates the median, the box indicates the inter-quartile range and the whiskers indicate the full range.}}
    \label{fig:magElev}
\end{figure}

Although it is the solar phase angle that determines the apparent brightness of a satellite for an observer, this is not very useful for astronomers. In figure~\ref{fig:magElev} we show the expected satellite brightness as a function of sun elevation angle, as seen by the observer and relative to the horizon. We see that the apparent brightness of Starlink satellites is insensitive to sun elevation angle through the day and reduces during the night, when the sun is below $\sim$15~degrees below the horizon. This explains why satellites are generally observed at twilight, when the sky is sufficiently dark but the satellites are still sufficiently bright, maximising the signal to noise ratio. The satellite brightness also has a dependence on the observation elevation angle due to the airmass dependent atmospheric extinction (see equation~\ref{eq:mag}).

\subsection{Apparent angular velocity of Starlink satellites}
\begin{figure}
    \centering
    \includegraphics[width=0.4\textwidth]{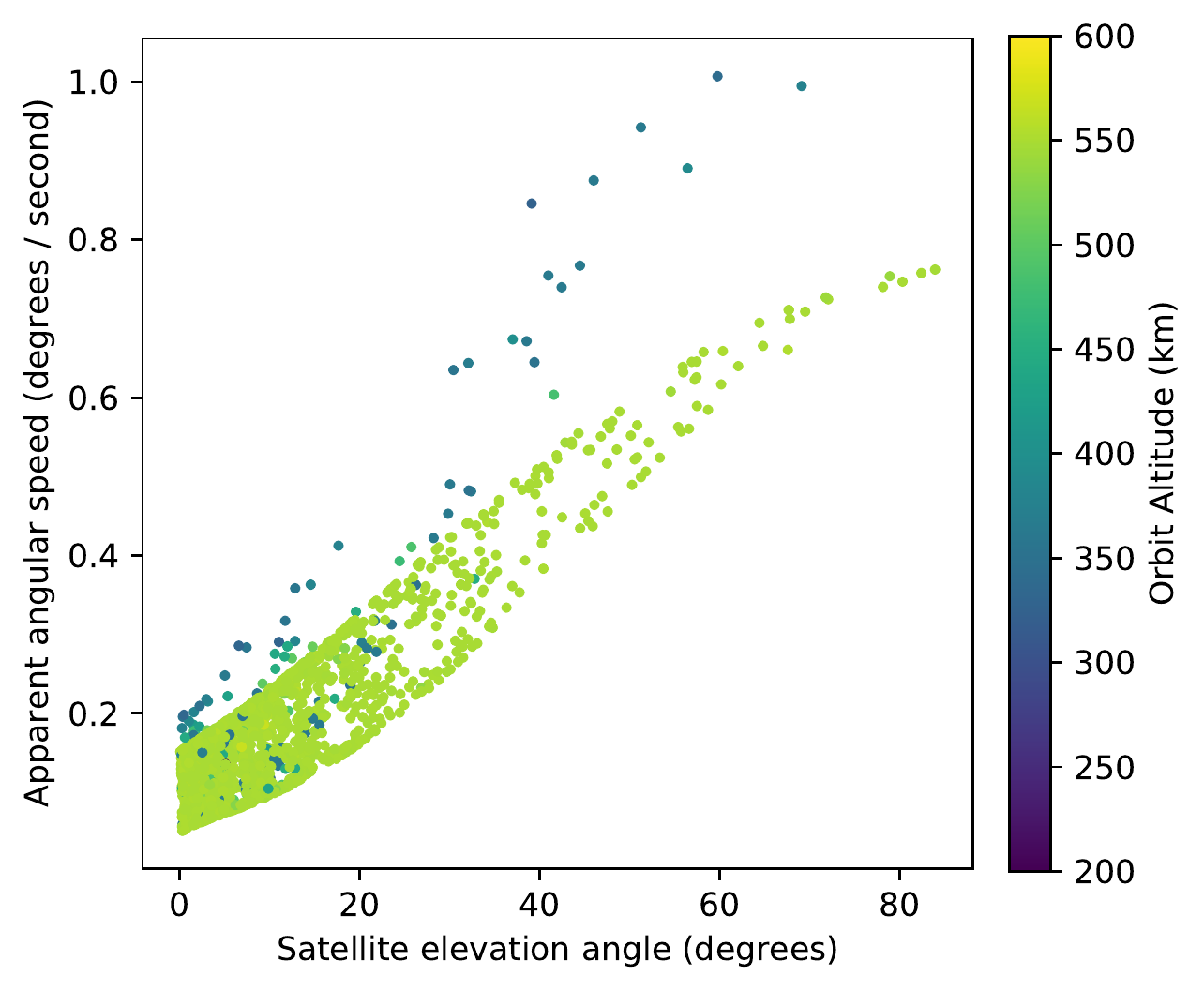}
    \caption{Angular speed of Starlink satellites as seen by an observer in a typical observatory at 25~degrees North. There are two distinct sets of objects visible. The majority of Starlink satellites follow a curve from approximately 0.1~degrees per second at the horizon up to approximately 0.8~degree per second at zenith. \changed{These correspond to the satellites orbiting at approximately 550~km. The scatter of the apparent angular speed is due to the projection effect determined by the relative direction of the satellite with respect to the position of the observer.
    There is another smaller set of faster satellites that correspond to lower orbits.
    }
    }
    \label{fig:satRate}
\end{figure}

In order to estimate the impact on astronomy we need to be able to assess the received flux per pixel per exposure. In addition to the number density and apparent brightness of the satellite, this requires some knowledge of the apparent velocity of the satellites. Figure~\ref{fig:satRate} shows the expected apparent angular velocity of the Starlink satellites as a function of elevation angle. At low elevation angles, projection effects lead to a reduced apparent angular velocity, with maximum angular velocity at zenith. The majority of Starlink satellites follow a curve from approximately 0.1~degrees per second at the horizon up to approximately 0.8~degrees per second at zenith. \changed{This main branch is attributed to satellites in operational orbits near to 550~km. Another branch can be seen corresponding to lower altitude and hence apparently faster satellites. These are satellites that are operating at lower altitudes, have recently been deployed into parking orbits, are in the process of orbit-raising, are in the process of de-orbiting or have failed.}
    
\section{Projection to the future}
There is considerable interest in the impact of future satellite constellations on astronomy and also on the naked eye visibility of the night sky. Using this model we can predict future satellite density and brightness based on projected constellation orbits and numbers. We define orbital parameters for a potential Starlink like constellation from the ASTRON package \citep{bassa2021analytical}.

\begin{figure*}
    \centering
    \includegraphics[height=5.5cm]{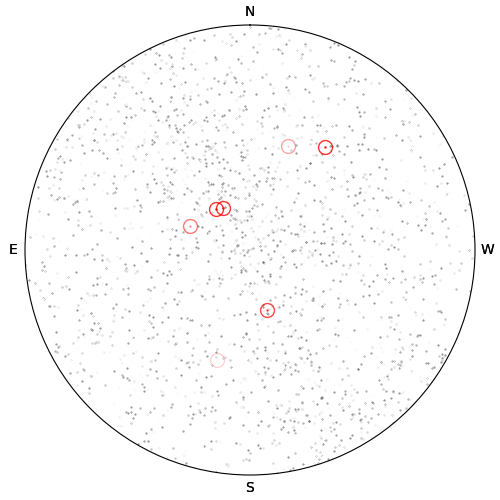}
    \includegraphics[height=5.5cm]{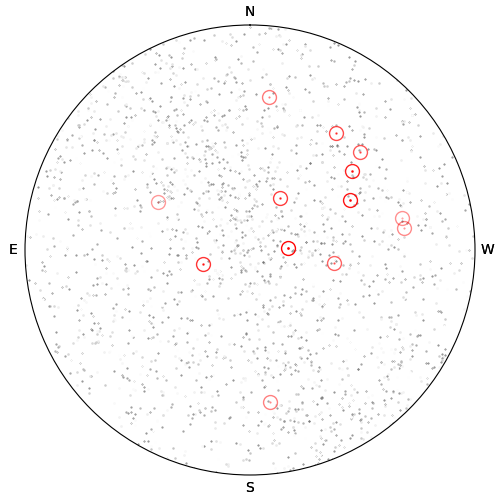}
    \includegraphics[height=5.5cm]{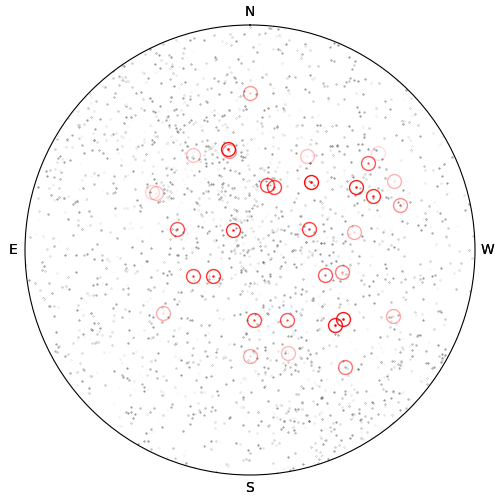}
    \caption{Example satellite visibility map down to magnitude \changed{6} in the V-band (naked eye limit at a very dark site) for an observer at 25~degrees North. The time is selected such that the sun is at 18~degrees below the horizon, \changed{classified as astronomically dark}. Stars are shown in black and satellites in red and are highlighted by circle to increase visibility. From left to right we show the naked eye visibility of an example LEO satellite constellation with 1600 satellites (left, current Starlink situation, \changed{7 visible satellites}), 12000 satellites (centre, approved Starlink plan by the US Federal Communications Commission, \changed{13 visible satellites}) and 40000 (right, planned Starlink constellation, \changed{31 visible satellites}).}
    \label{fig:future_visible}
\end{figure*}

Figure~\ref{fig:future_visible} shows examples of the current and future maps of naked eye visible satellites at a \changed{very} dark site (<magnitude \changed{6}) \changed{at the end of astronomical twilight (sun is 18~degrees below the horizon).} 
\changed{
This is a worst-case model.
This is only valid for the very best sites with no light pollution and observers with very good eyesight.  
Also, later in the night there is a lack of bright satellites due to the observer - satellite - sun geometry and earlier in the night the increased sky brightness will reduce the brightness limit meaning fewer satellites can be seen.
}

There is no difference in the relative brightness distributions but the numbers of visible satellites increases dramatically\comment{ (Figure~\ref{fig:visSatHist})}. The figures shows that the expected number of visible satellites during twilight (ie the maximum number) goes from approximately \changed{5 for 1600 satellites, to 10 and 30} for 12000 and 40000 satellites respectively. 

This does rely on our brightness model being statistically accurate.\comment{ Although there is some support for this for existing Starlink satellites \citep{Hainaut2020}, it should be noted that SpaceX are actively engaging with astronomers and there are ongoing attempts to reduce the apparent brightness as seen from the ground \citep{Tyson2020}. Recent studies suggest that the new design of Starlink satellites could be reduced by a factor of 2 \citep{Horiuchi2020} (almost one magnitude).}

\comment{
\begin{figure}
    \centering
    \includegraphics[width=0.4\textwidth]{figures/vissathist.pdf}
    \caption{Histogram of satellite brightness for a LEO constellation of 1635 (blue), 12000 (orange) and 40000 (green) satellites visible during twilight. The dashed line indicates the limit of naked eye visibility.}
    \label{fig:visSatHist}
\end{figure}
}
\section{Conclusions}

We present a new software tool, Astrosat, written in python and available at https://github.com/james-m-osborn/astrosat. Astrosat is based on a central API which can be easily scripted into stand-alone tools or integrated into other software pipelines. 

Astrosat allows users to probe expected satellite / space object transit events through a given field of view for any observer location, time, field of view and observation duration. Astrosat can be used as an API which can be integrated into existing pipelines or as a standalone tool. We expect astronomers will be able to use the tool to schedule observations to minimise satellite impact, to shutter transit events or to identify artefacts in data by using the expected transits as a prior in data analysis pipelines. In addition, landscape and astro-photographers can  use the tool to plan their activities in order to avoid or even include the satellite trails depending on their desired impact. The tool can also be used for outreach activities to visualise and demonstrate the effect of the satellites on scientific observations but also how our naked-eye view of the night-sky is being impacted.

We have presented basic analysis such that astronomers can assess the expected impact of Starlink and similar satellite constellations. We have shown the expected number, apparent brightness and angular speed of the satellites all as a function of elevation angle. 

As expected, more satellites can be seen at lower elevation angles due to the projection effect. For the same reason, these satellites also appear to move more slowly ($\sim$0.1 degrees per second near the horizon compared to $\sim$0.8 degrees per second at zenith).

Interestingly satellites tend to have a consistent brightness during the daytime and become fainter at night as they move into the shadow of the Earth. However, they are most visible during twilight when they still have the same apparent brightness but the background from the sky is significantly reduced. The brighter satellites tend to be seen at higher zenith angles and occur at twilight when the angle subtended between the sun, satellite and observer is optimal.

In addition we also show example visualisations of the night sky at twilight for 1600 (current number of Starlink satellites), 12000 (number of authorised Starlink satellites) and 40000 (number of planned Starlink satellites). We expect a naked eye observer will be able to see \changed{ a maximum of} approximately \changed{5, 10 and 30} satellites during the twilight period for each scenario respectively.

\section*{Acknowledgements}
JO acknowledges the UK Research and Innovation Future Leaders Fellowship (MR/S035338/1) and the Science and Technology Facilities Council (STFC) through grant ST/P000541/1. 

\changed{This research has made use of the SIMBAD database, operated at CDS, Strasbourg, France.}

The simulations in this paper make use of the NumPy \citep{numpy}, Scipy \citep{scipy}, AOtools \citep{Townson2019}, and matplotlib \citep{matplotlib} Python packages.

\changed{We would like to thank the reviewer, Dr. Hainaut, whose insightful comments certainly improved this paper and helped to make astrosat more complete.}

\section*{Data Availability}
No data was archived for this publication. Data for plots can be generated using Astrosat.



\bibliographystyle{mnras}
\bibliography{astroSat} 



\comment{

\appendix

\section{Some extra material}

If you want to present additional material which would interrupt the flow of the main paper,
it can be placed in an Appendix which appears after the list of references.

}

\bsp	
\label{lastpage}
\end{document}